\documentclass[preprint]{revtex4}  
\usepackage{CJK}   
\usepackage{indentfirst}  
\usepackage{bm}    
\usepackage{graphicx}  
\usepackage{caption}
\usepackage{booktabs}
\usepackage{supertabular}
\usepackage{ulem}
\usepackage{color}
\usepackage{dcolumn}
\begin{document}    

\title{On the energy conservation  electrostatic PIC/MC simulating: benchmark and application to the radio frequency discharges \footnote{ to be appeared in Chinese  Physics   B}}  

\author{Wang Hong-Yu$^{1}$} 
\author{Sun Peng$^{1}$}
\author{Jiang Wei$^2$}
\author{Kong Ling-Bao$^{3}$}

\affiliation{
${}^{\rm 1)}$  School of Physics Science and Technology, Anshan Normal University, Anshan, 114005, China \\
${}^{\rm 2)}$ School of Physics, Huazhong University of Science and Technology, Wuhan, 430074, China \\
${}^{\rm 3)}$ School of Science $\&$ Beijing Key Laboratory of Environmentally Harmful Chemicals Assessment, Beijing University of Chemical Technology, Beijing 100029, China }
\date{\today}

\begin{abstract}
We benchmark and analyze the error of energy conservation (EC) scheme for Particle in cell/Monte-Carlo Couple (PIC/MCC) algorithms by a radio frequency discharging simulation.
The plasma heating behaviors and electron distributing functions obtained by 1D simulation are analyzed. Both explicit and implicit
algorithms are checked. The results showed that the EC scheme can eliminated the self-heating with wide grid spacing in both
cases with a small reduction of the accuracies. In typical parameters, the EC implicit scheme has higher precision than EC explicit scheme.
 Some "numerical cooling" behaviors are observed and analyzed. Some other error are analyzed also.  The analysis showed
 EC implicit scheme can be used to qualitative estimation of some discharge problems with much less computational resource costs without much loss of accuracies..

 \end{abstract}

 \pacs{ 52.80.Pi , 52.27.Aj, 52.65.Rr}
 \keywords{Particle in Cell / Monte Carlo Couple, Energy Conservation, Grid Heating, Discharging simulation}

\maketitle

\section{Introduction}  

article-in-cell (PIC) simulation is a key method for tenuous plasma modeling{\cite{Birdsall91}}. In PIC simulation, the kinetics effects are considered self-consistently. On the other hand, PIC simulation gets rid of the full modeling in phase space through the particle sampling{\cite{Verboncoeur05}}. With Monte--Carlo Couple (MCC) method, it can be used to simulate the discharge problems in detail. As a result, PIC simulation becomes an important
technology in dealing with the problems of gas pressure discharge{\cite{Lieberman05}\cite{Zhao08}\cite{Shi09}\cite{Liu10}\cite{Jin09}\cite{Wang13}}.
However, PIC simulation has some shortcomings{\cite{Tskhakaya07}}. One of them is the space--time scale limit. In convenient PIC algorithms, the grid spacing and
time steps are constrained by

\begin{eqnarray}
\nonumber \Delta t< \frac {2}{\omega_p},\\
\nonumber \Delta t <\frac {\Delta x} c,\\
\Delta x<\xi \lambda_D,
\end{eqnarray}
where $\xi$ is a factor having order of 1  and is determined by the electron velocity distributions. For electrons with Maxwellian distribution, it is  $\xi\sim 3$. These constraints come from the leap-frog algorithms,  Courant--Friedrichs--Lewy limit (CFL) and fine grid instability (FGI) separately{\cite{Tskhakaya07}{\cite{Hockney88}}.
In discharge plasma devices, these constraints cause very large number of the grid cells and the time steps, so that the simulation can
cause very large computational costs.

In the typical low temperature discharging plasmas, the unpractical huge computational costs cause severely difficult problems. It is because that both the leap-frog limit and the fine grid instabilities become critical (even the FGI can be neglect in many other cases). When the leap-frog constraint is violated, the simulating diverges immediately. When the FGI constrain is violated, an effect named``self heating" takes place and increases the particle kinetic energy. However, in the gas discharges, the plasma is kept by collision and the collisional cross sections are sensitive to the particle energy. When the self-heating tries to increase the particle energy,
more collisional ionizing should occur instead. Hence, the self-heating does not increase the energy of particles largely but increase the densities of plasma.
Furthermore, the increasing plasma densities decrease the Debye length and cause more self-heating. This positive feedback intensively disturbs the simulating and leads to a slow diverge often. Therefore, in the discharge PIC/MCC simulating, we need to use very small time and space steps. Because of the
computational resource costs being proportional to $1/(\Delta x^{D} \Delta t)$ in D-dimension simulating, it becomes impractical huge in the industrial applications. This difficulty is called``space--time scale problem". In multi-dimension simulating, the FGI constraint is more critical.

To overcome the space--time scale problem, we can apply some variants to the algorithms or the physical modeling. For example, the scaling method
resizes the simulation region by adjusting the colliding rates to reduce the computational costs{\cite{Taccogn04}}. This method depends on the scale similar law and
can be used in very narrow regions. We should mention that there are other tricks to mitigate the grid self-heating (sometimes is called ``grid heating''). The method is to apply the high-order weighting scheme to the particle/current and field weighting. This trick can reduce the aliasing error greatly and slow the increasing of self-heating{\cite{Abe86}}. Hence, it has been applied to solve the problems when the simulating time is not extremely long\cite{Sentoku08}\cite{Cai10}. However, the self-heating is slowed down instead of being eliminated. In addition, very high order weighting scheme will introduce complexity in parallelization.

There are also two kinds variant algorithms overcoming the space-time limits. One is implicit PIC method {\cite{Langdon83}\cite{Friedman90}} and the other is energy-conservation (EC) algorithm{\cite{Birdsall91}{\cite{Hockney88}}}. Implicit PIC method eliminates the time step limits  by damping the high-frequency oscillations in time domain, consequently, it will be stable for very long $\Delta t$. The damping also decreases the self-heating of the plasma. When the time step is increased by the grid spacing, the self-heating will be decreased and FGI will be eliminated{\cite{Wang10}}.

The EC scheme introduces the total energy (field energy plus particle kinetic energy) conservation of the plasma. Theoretically, the self heating takes place in some modes and increases the particles kinetic energy. When the total energy is constrained to conserve, the self heating can not increase to diverge, then the EC scheme can eliminate the FGI constraint but does not affect the time step limit.

Both of the two algorithms are used in large-scale or high density plasma simulations. Implicit algorithms are used in  capacitive coupled plasma (CCP) discharging{\cite{Vahedi93}}, geophysics{\cite{Ricci02}}, and high energy densities physics{\cite{Welch06}}. Some variants of EC scheme are used in some electromagnetic PIC simulating softwares{\cite{Eastwood91}\cite{Eastwood95}}.  One-order weighting EC scheme has some problems as stochastic noise. However, high-order weighting can cure them. Recently, some development of EC is introduced to simulate the high-power devices{\cite{Pointon08}}. We should point out that these EC schemes are always
approximate energy conservation with finite d$t$. One kind  of accuracy energy conservation PIC algorithms are introduced {\cite{Markidis11} and   promise well working in the future. However, these algorithms are very complex and difficult in the equation solving and coding, which we will discuss in another paper.

The reduction of computational resource causes the reduction of accuracy. e.g., implicit PIC simulating can cause plasma self-cooling{\cite{Brackbill85}}. We can recognize that the implicit PIC method worked by the whole region oscillation damping, and the grid heating source is not cured indeed because the grid heating does not relate to the Langmuir oscillation.  Similarly, the EC scheme suppresses the
energy increasing from the grid heating. However, some self-force is introduced and the particle distribution in phase space can be disturbed{\cite{Langdon73}}. Therefore, it
could cause some errors of kinetic behaviors.

Therefore, we need detail benchmarks and errors analysis of these methods' application in discharge simulation. Vahedi{\cite{Vahedi93}} studied the implicit methods and compared them with the explicit simulation in CCP, while no benchmarks are performed for the EC algorithms with MCC up till now. Hence, the detail investigation of the accuracy and errors in EC algorithms is needed. In addition, the effective combination of EC scheme and implicit algorithms has not been reported.

In the present paper, we analyze the simulation results and compare convenient PIC algorithms (referred as``momentum conservation scheme" or ``MC sheme") with  EC algorithms and obtain the validity and applicability of these algorithms. The benchmark model is a one-dimensional CCP discharging plasma. Because the driving frequency is as low as tens of MHz, electrostatic modeling is used. To model the collision in plasma, MCC procedure is applied. We show that the combined EC-implicit PIC algorithm
could be a qualitative method for the discharge simulating. In addition, the errors and the origin of its errors are discussed.

\section{the numerical algorithms of PIC and EC scheme}  

In this paper, we test and analyze a series PIC algorithms. The base algorithms are the convenient explicit PIC algorithms ("explicit") and the direct implicit PIC algorithms ("implicit"). The detail of these algorithms can be found in our early papers{\cite{Wang10}\cite{Jiang08}}. In the early paper, all
of the algorithms used momentum conservation scheme(MC). At the present paper, we patch them with the energy conservation scheme (EC). The major idea of EC scheme can be found in {\cite{Pointon08}}. We describe it here very briefly.

The energy conservation scheme has only slightly different from the momentum convenient conservation scheme. Denote the middle point of $x_g$ and $x_{g+1}$ by $x_{g+1/2}$,
then the weighting of the charge should be alternated to
\begin{eqnarray}
\label{weighting}
\rho(x_g)&=&\sum_p q_pS_m(x_g-x_p)
\end{eqnarray}

The interpolation of fields should be alternate to
\begin{eqnarray}
\label{interpolation}
F(x_p)&=&\sum_g qE_gS_{m-1}(x_p-x_{g+1/2})
\end{eqnarray}

Where $S_m(x)$ is the standard PIC interpolation function (see {\cite{Abe86}}).
The other algorithms are not changed(Both for explicit and implicit algorithm). Then there are four PIC schemes:

(1) MC explicit scheme, the scheme is same as ref \cite{Jiang08}. In fact, this is the standard electrostatic Particle-in-cell/Monte--Carlo
Algorithms in literature.

(2) EC explicit, the scheme applies the patch to  the ref \cite{Jiang08}.

(3) MC implicit,  the scheme is same as ref \cite{Wang10}. In literature, this algorithms are called "Direct Implicit Particle-in-cell"
method.

(4) EC implicit, the scheme applies the patch to the ref \cite{Wang10}.

In consideration of reduction stochastic noise by interpolation, one should set $m=2$ and larger $m$
 can be applied also. In this paper, we choose $m=2$ in the EC scheme only. Under this chosen, only the
 particle weighting (equation {\ref{weighting}}) become more complex and the field interpolation is almost same as the
 convenient scheme. Then the coding work and the computational cost increasing is minimum.

Theoretically, the simulation running time equals $t=T_{physical}/{\Delta t}*t_{step}$. Where $T_{physical}$ is the physical time of the system and $t_{step}$ is the time for one PIC step simulating. Then increasing $\Delta t$ will decrease the simulating costs but the increasing is limited by leap-frog constrains and MCC collision. $t_{step}$ is related to the field solving time $t_f$ and the particle pushing/monte-carlo time $t_{ptcls}$. For the Monte-Carlo subroutine's accuracy and noise controlling, the PPC must be set to larger than some limit(say, 100).  If we fix the particle number per cell (PPC) according to  the monte-carlo code required, $t_{ptcls}$ will be in direct proportion to the cell number and $t_f$ will also be positively correlated to the cell number. Thus increasing the grid spacing will reduce the computational cost. On other hand, our EC scheme needs 2 order weighting of particles, while implicit algorithms need some more complex pushing code. Hence EC and implicit scheme will run slightly slower than convenient PIC scheme when same grid spacing is applied. Anyway, in almost all cases, the grid spacing effects are dominant.

\section{The discharge model and the simulation}
We simulated a CCP discharging with above 1D model. In the CCP discharges, the electrode are driven by radio-frequency (RF). The electrons bounce on the sheaths and get slight acceleration. Then the electrons are scattered by gas atoms in the background and the electrons velocities are de-phased. In this processes, the energy of electric field is converted into the electron stochastic
energy. The energy conversion is called stochastic heating. The stochastic heating transfers energy into the plasma and causes the ionization. This mechanism is the  dominant physics issue of low gas pressure CCP.
This behavior exists widely in many kinds of low pressure discharge problems. Similarly, radio-frequency  sheath formation and stochastic heating are key mechanisms of these discharges. Therefore the analysis and benchmark of them in CCP simulation can be used to estimate the
effects of the algorithm applying to these discharge problems.

The geometry of the model is of 1D symmetric planar for the sake of simplicity. The Ar gas
is put between two electrodes with radio-frequency voltage source being applied.
The frequency of RF source is 27 MHZ and the voltage amplitude is 200V.
The gap length between electrodes is $4 cm$.
MCC procedure identical to ref{\cite{Jiang08}} is used to describe the electron-atom and ion-atom collisions.
The gas pressure is set to 50 mTorr and the secondary electrons are neglected.

The conceptual behavior of the CCP simulating can be described by energy balance. The heating process is the discharging plasma generator. On the other hand, when electrons and ions diffuse to the electrode, they are absorbed.
This process is the plasma eliminator. At the beginning of the simulation, the plasma are filled uniformly between the electrodes. Then
the bipolar diffusion causes the sheathes formation. After the sheathes formation, the stochastic heating and the electrode absorbing compete with each other and evolve the plasma densities. When the two effects reach a balance, the simulating converge to a stable results.

We can recognize that if the grid heating is too large, the simulating will diverge for the positive feedback mentioned in section I. However,
the absorbing rates is related to the plasma densities. If the grid heating is not very large, the simulating can still converge. Because of the error of the total heating(and the total heating rates  profile), the simulating results are essentially wrong. In addition, when the gas pressure
is  low, the stochastic heating causes the bi-Maxwellian while the colliding and absorbing cause cutting Maxwellian distribution. Hence, even if the simulating
converges, the simulating can possibly be wrong without careful check. On the other hand, the complexity and the sensitivity of CCP discharging causes it
to be a good tester of the PIC/MCC algorithms.

In our benchmark simulations, the initial plasma densities are set to $2.0*10^{16}/m^3$ uniformly and are lower than the densities when reaching equilibrium. With the simulation running, the macro particle numbers and the plasma densities will increase slowly(after a short decrease to construct the shealth).
If the simulating converge, the macro particles number will converge to a fixed value, when all physical parameters are given, and will only fluctuate slightly around it.
In
the convenient PIC regime, we set the grid number to 800 and could see the simulating converge after about $16000-22000$ RF periods. Theoretically,
for the electron Maxwell distribution, when the $DX<3\lambda_D$, the grid heating effects can be ignored\cite{Tskhakaya07}. In this case, the stable plasma density is about $5.2*10^{16}$, and the electron Debye length is about $3.3*10^{-5} m$. Thus the $dx \sim 1.4 \lambda_D$ and $\omega_p dt \simeq 0.2$. When the grid spacing increases to more than 3 $\lambda_D$ , the macro particles number will keep increasing and diverge finally. Similarly, in the momentum conservation direct implicit
PIC simulating, the diverge can be observed too. The phenomena take place while the time step $dt$ of implicit simulating is small and the grid spacing is much larger than $\lambda_D$ . In contrast, the EC scheme causes converge results after likely RF periods for all the grid spacing but the stable plasma densities are not same for the different grid spacings. When the grid spacing becomes lower than $\lambda_D$ and the time step $dt<0.2/\omega_p$, all of the codes get equal densities as expected.

As the discussion in section 2, the increasing grid spacing can greatly reduce computational costs . For the converge periods in different cases are close, we can
benchmark the computational time per RF period ($T_{period}$) to estimate the simulation costs. Table {\ref{time}} give a simple benchmark results on a PC  with $PPC=400$  in all cases.

\begin{table}[htbp]
\caption{$T_{period}$ by second with alternative algorithms and grid spacing. We choice a status with same physical time in the simulation, and $dt=1.5*10^{-11}$ for all cases. 'N/A' means this simulation will not converge.}
\label{time}
\begin{ruledtabular}
\begin{tabular}{ r  c c c c c r }
grid Number & MC explicit & EC explicit & MC implicit & EC implicit \\
\hline
 800 & 32.5 & 35.2 & 41.7 & 57.5  \\
  400 &15.9 & 16.7 & 21.0 & 26.5\\
  200 & N/A & 8.0 & N/A & 12.3 \\
   64 & N/A & 2.3& N/A & 3.1\\
 \end{tabular}
 \end{ruledtabular}
 \end{table}

One can recognize that the computational costs can be reduced one order in 1D cases with increasing of  the grid spacing when one applies the EC scheme: At the benchmark system, MC explicit grid=800 simulating runs about 10 days while EC implicit grid=64 simulating runs less than one day.
If the simulating is 2D modeled or the time step $\Delta t$ is increased, the costs will be reduced much more( see as \cite{Wang10}).

Anyway, the macro particle number can affect the stochastic noise and cause error in the results. In most of simulations, we set
initial macro particle number per cell (PPC) to 400 except the section about the PPC effects.  We will discuss the effects of these parameters as follows.

\subsection{The effects of grid spacing}

We should point out that the grid spacing limit is more critical than the time step limit in discharge simulating. The reason is the grid number increasing as ${dx}^{-D}$ where $D=1,2,3$ is the simulation's dimension. Therefore if one increases $dx$ 10 times, the resource costs decreases 1000 times for 3D simulating. In addition, the CFL can relate the time step to the grid spacing:
 $v_{max}\Delta t<\Delta x$，where $v_{max}$ is the max physical characteristic velocity in the systems. Then the $\Delta t$ could be increased while the $\Delta x$ is increased and the computational cost will be reduced more.
 In fact,  if the grid spacing can be increased greatly and the convergence can be kept, one can endure some system error of the simulation. Here we will discuss the grid spacing effects of the EC
scheme.

We believe that when the grid becomes dense enough, the results from different schemes will become same. In this case, when $grid=800$ ($DX=1.4\lambda_D$), the four schemes give very close
results and can not be distinguished from each other in the figure. More dense grid causes little change.  Hence, we use the MC explicit results with 800 cells to a benchmark data. I.e., the closer a result is to MC explicit 800 result,  the higher accuracy is .

\subsubsection{the plasma densities}

We show the plasma densities and average electric potential with stable discharging in figure \ref{eleden}, figure \ref{ionden} and figure \ref{Phi}.
The time step is set to $1.5\times 10^{-11}$ seconds and the grid numbers vary from 800 to 64. Because we are interested in the EC scheme, we showed the
MC explicit results for benchmark only.

For all cases, the potential profiles are very similar and the values are close. Difference of the potential is only observable when the grid spacing is too large. The plasma densities seems to be more sensitive to the grid spacing. As we known, the MC explicit PIC simulation will diverge when the grid spacing is larger than $3\lambda_D$.

When the grid spacing increases, the EC scheme will give lower plasma densities. Even when the grid spacing becomes much larger than Debye length, the profile of plasma densities are still similar and can be fitted with a multiplying factor. In the most extreme cases, when $\Delta x\sim 18\lambda_D$,
 the plasma densities are equal to about 60\% of the $\Delta x= 1.4 \lambda_D$ cases by the EC explicit scheme. It seems that the EC implicit scheme gets more accurate results: When grid number is equal to 64, the result error is less than 20\%. Because of the converge characteristic of EC PIC algorithm, we think it can be used to estimate qualitatively the plasma distribution for the engineering. If the time step constraint is not very critical, both the EC explicit scheme and EC implicit scheme can be applied to qualitatively analysis. However, for the very wide grid, the EC implicit scheme can work better.

\subsubsection{Sheath behavior}
For the CCP discharge, the sheath behaviors are very important. In fact, many of the interesting processes and the major physical mechanism are
working in the sheathes. For the ratio frequency CCP, the sheathes are oscillating. We show the average $\delta n=n_i-n_e$ in figue \ref{sheath}. The profile of the $\delta n$ can be seen as the sheath boundary. One can find that the sheath thickness is all the same. Especially, when grid=64, the EC Implicit scheme can give out correct potential while EC Explicit scheme show about 5\% error on the potential. Because the sheath thickness is determined by the particle diffusion, we conclude that the EC scheme can model the particle diffusion process correctly.

\begin{figure}

\includegraphics[scale=0.4]{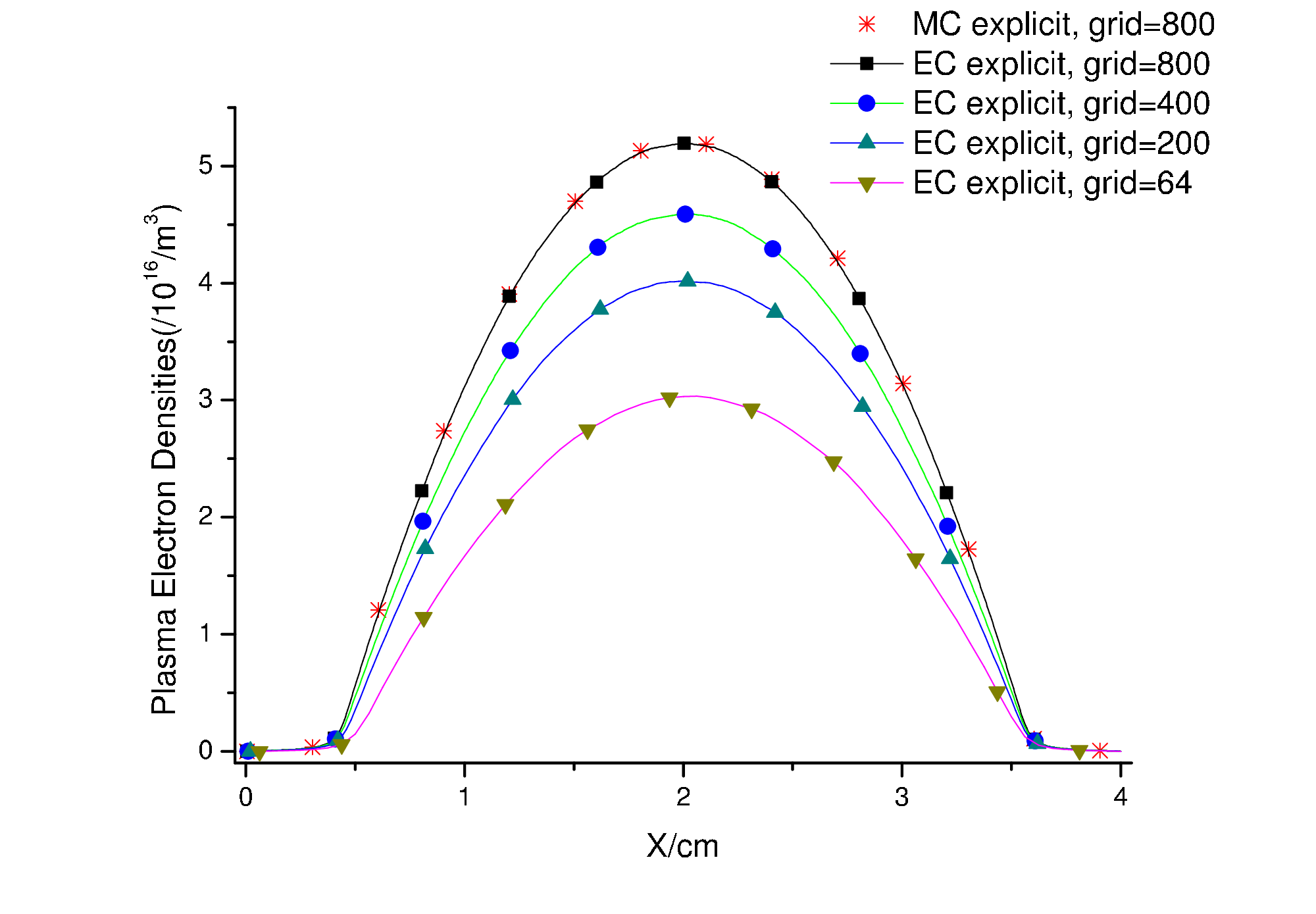}
\includegraphics[scale=0.4]{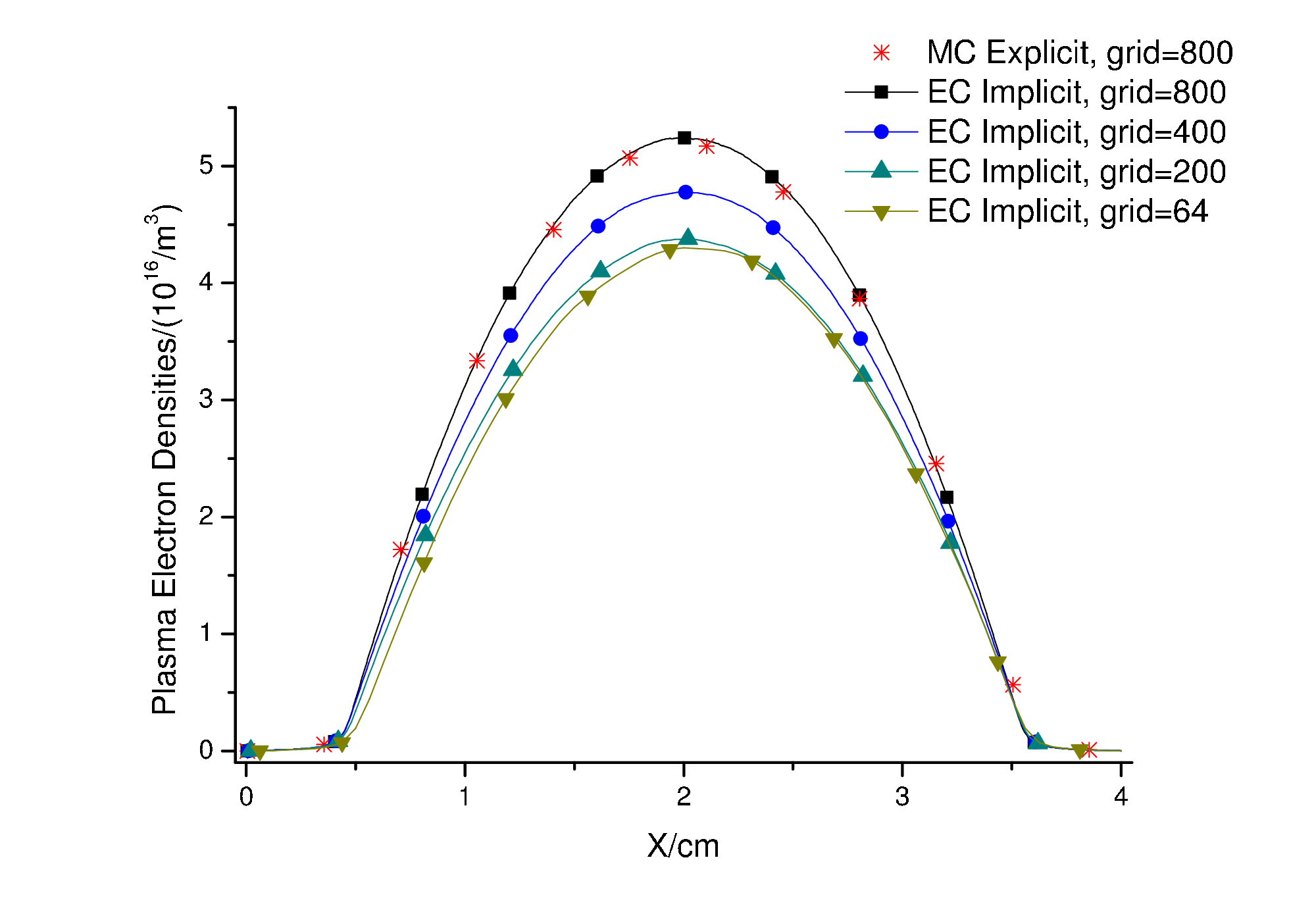}
\caption{Electron densities by EC scheme, left for Explicit, right for Implicit}
\label{eleden}
\end{figure}

\begin{figure}
\includegraphics[scale=0.4]{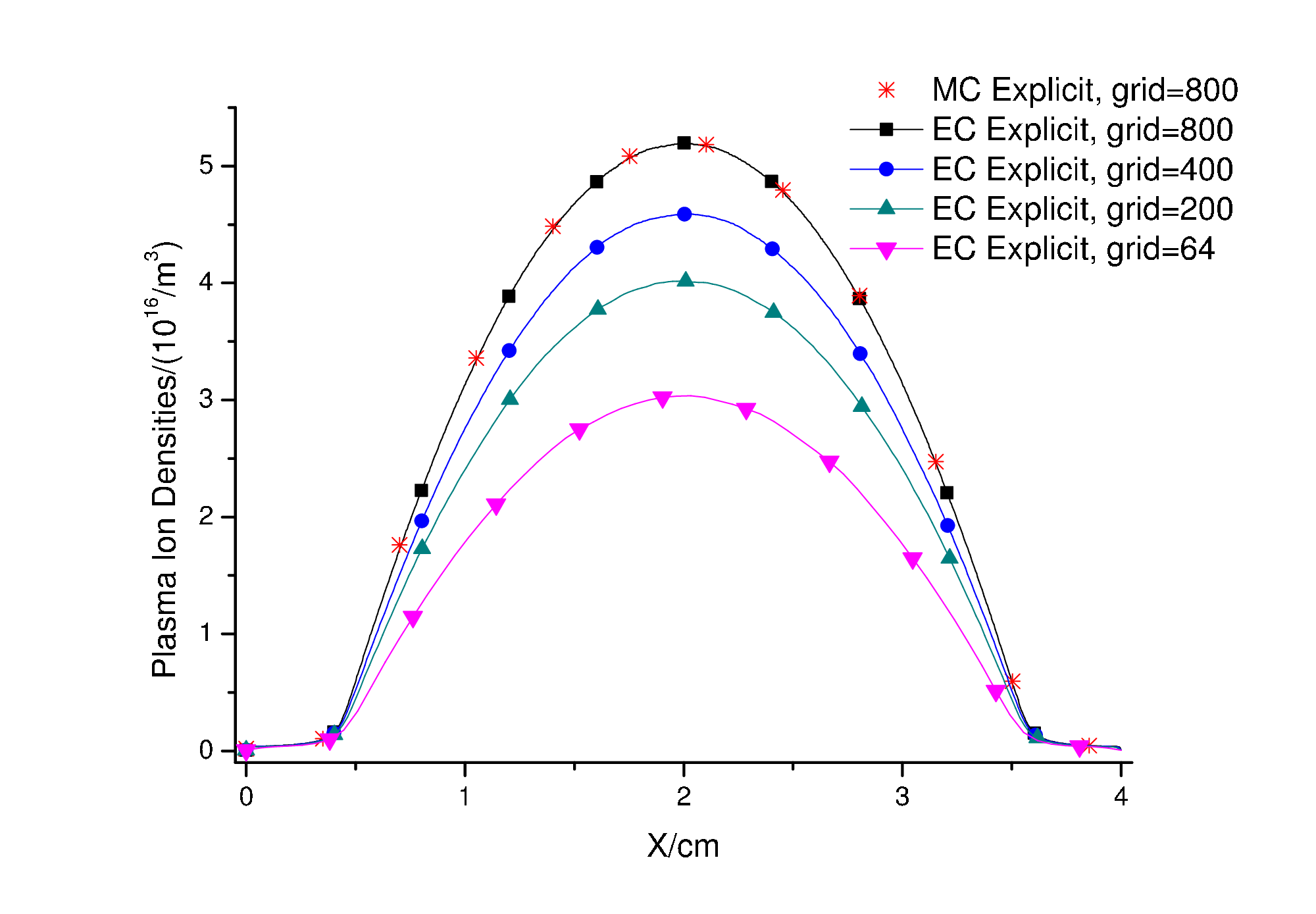}
\includegraphics[scale=0.4]{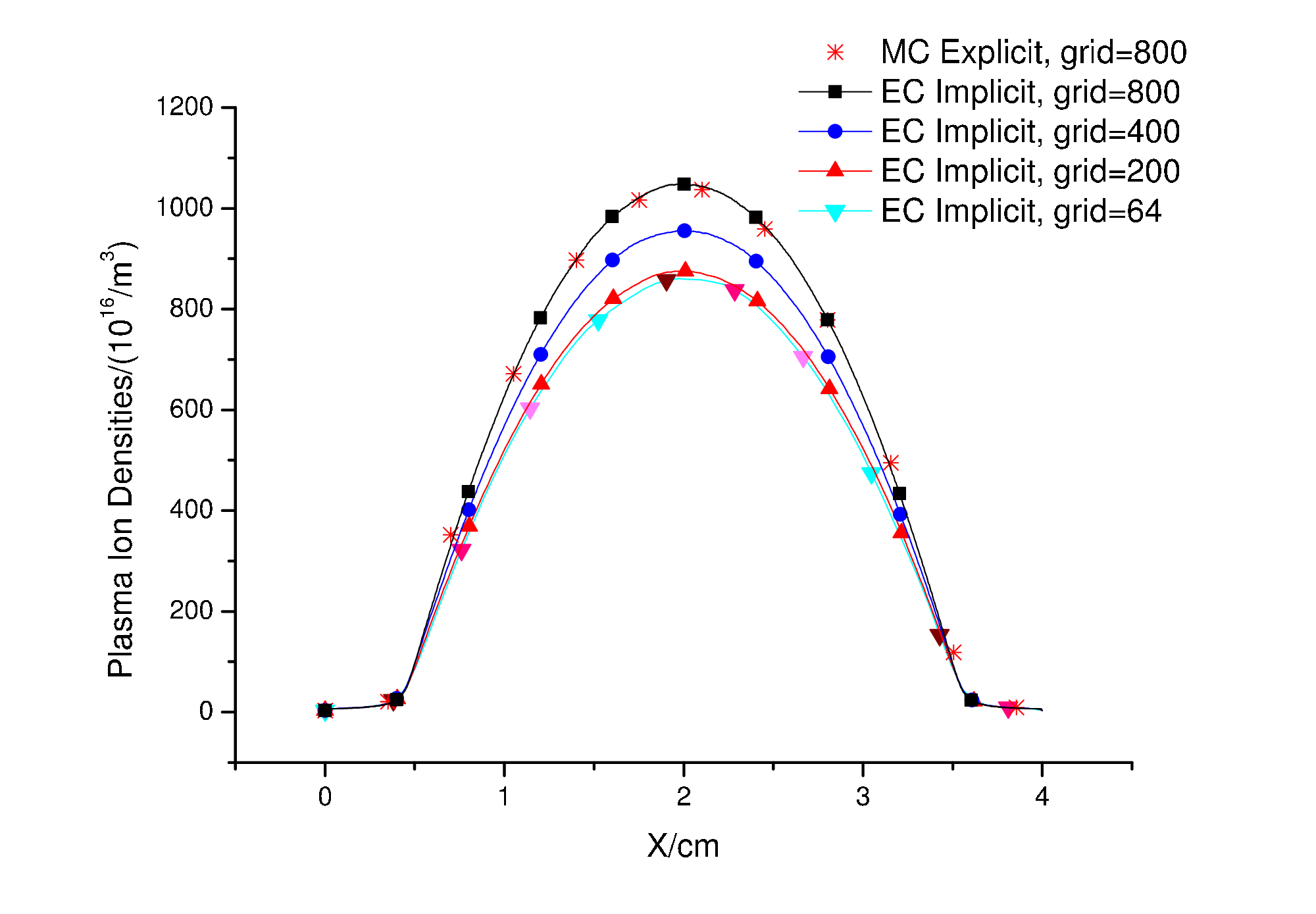}
\caption{Ion densities by  EC scheme, left for Explicit, right for Implicit }
\label{ionden}
\end{figure}

\begin{figure}
\includegraphics[scale=0.4]{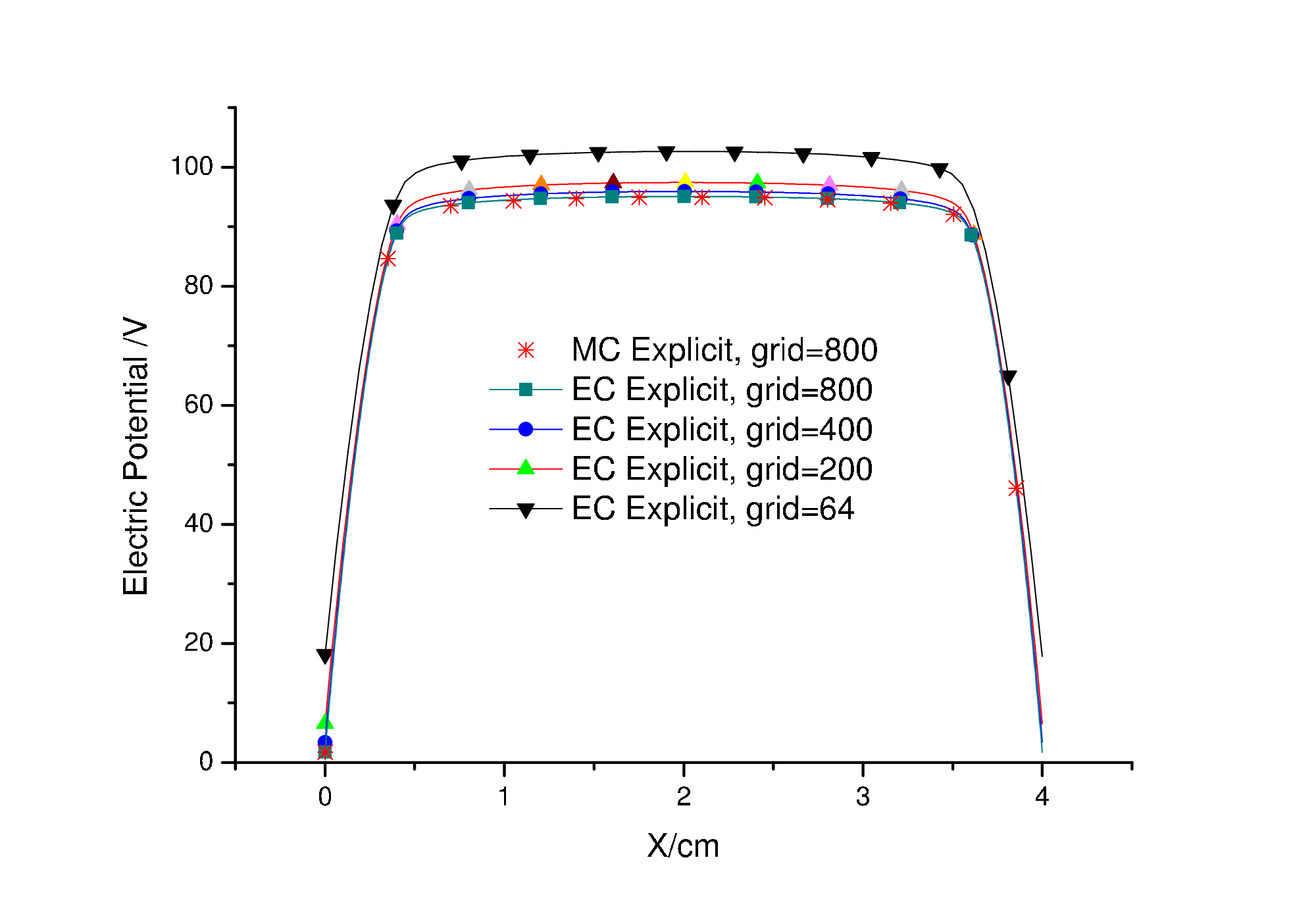}
\includegraphics[scale=0.4]{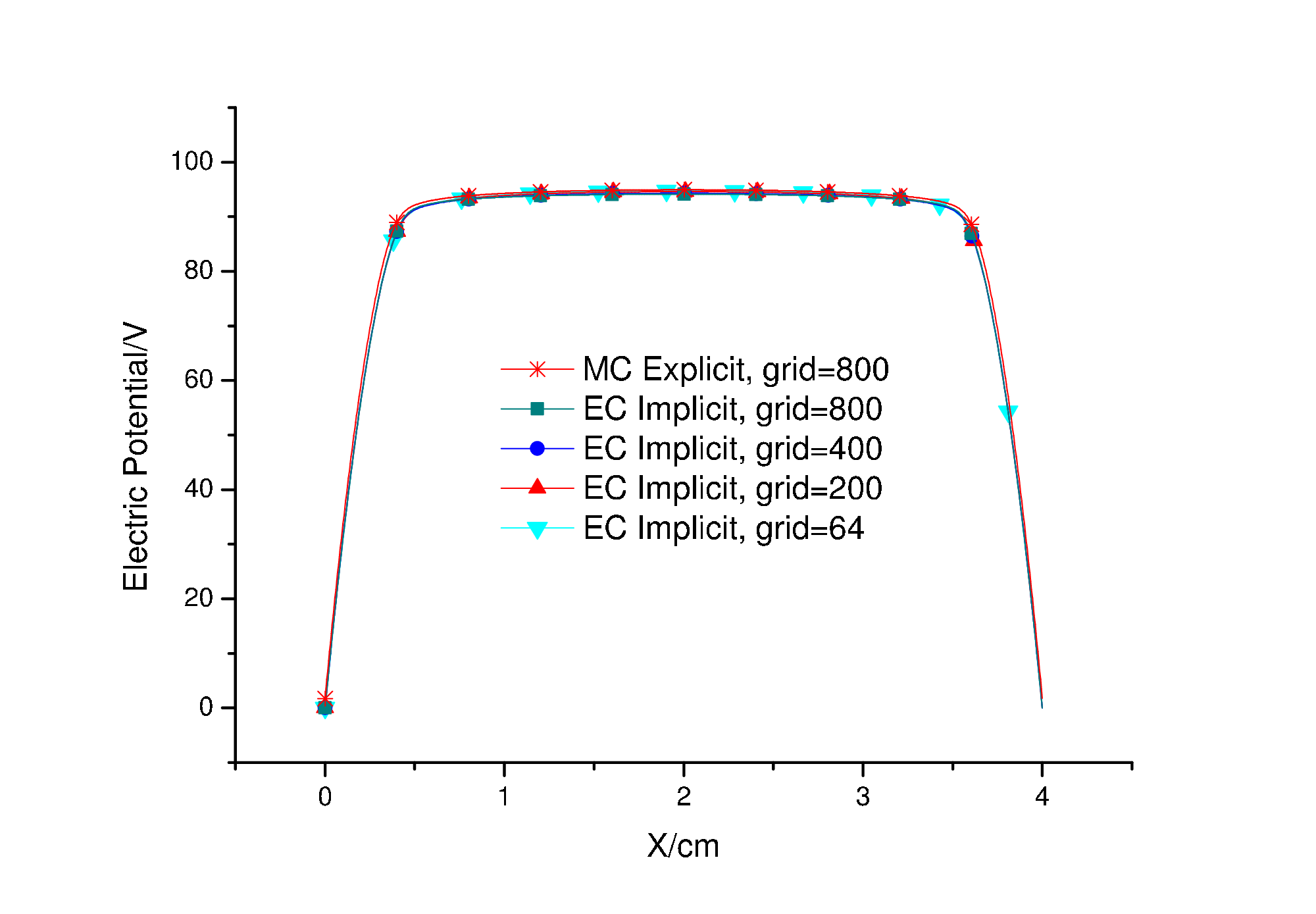}
\caption{Average Phi, left for Explicit, left for Explicit, right for Implicit }
\label{Phi}
\end{figure}

\begin{figure}

\includegraphics[scale=0.4]{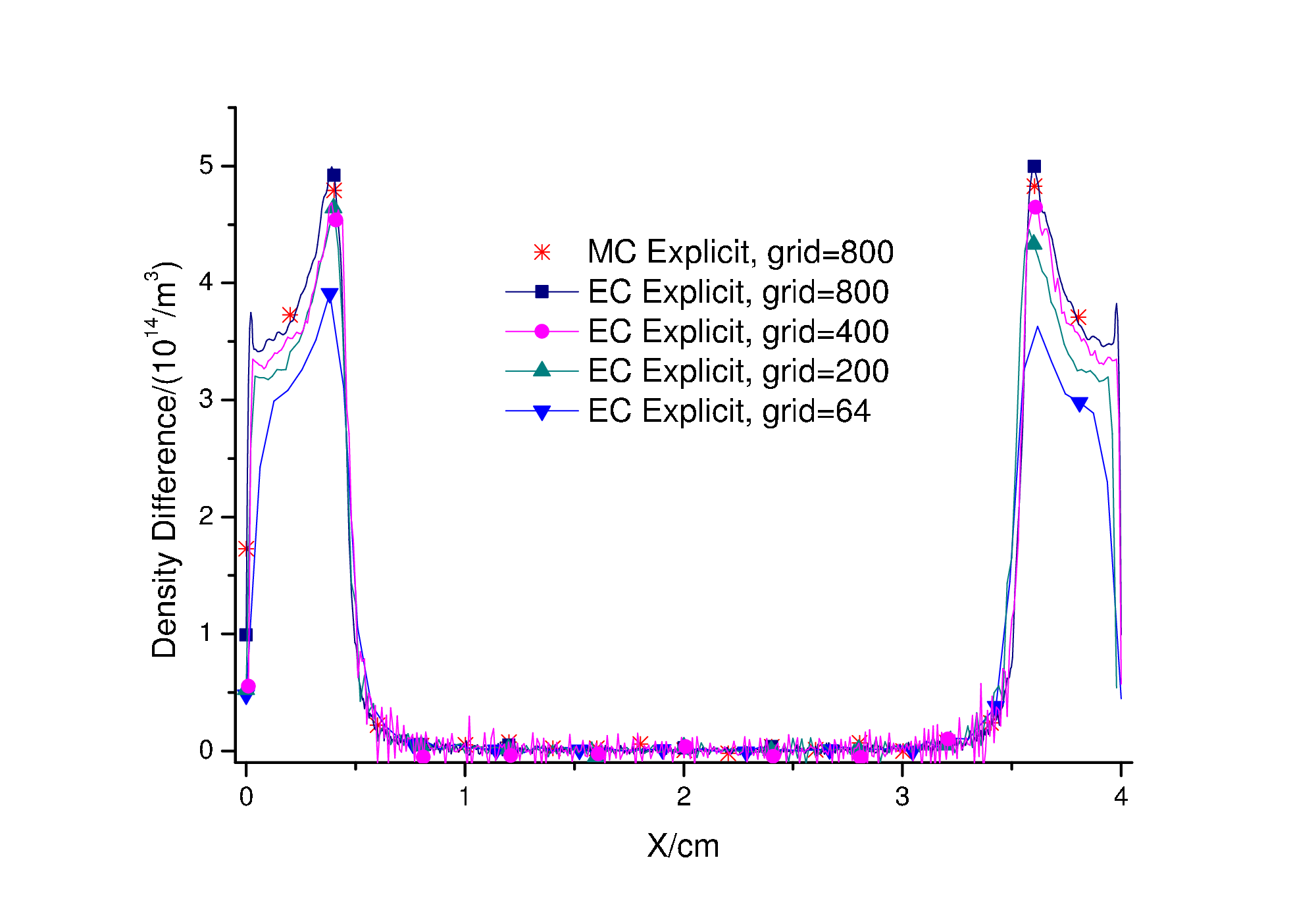}
\includegraphics[scale=0.4]{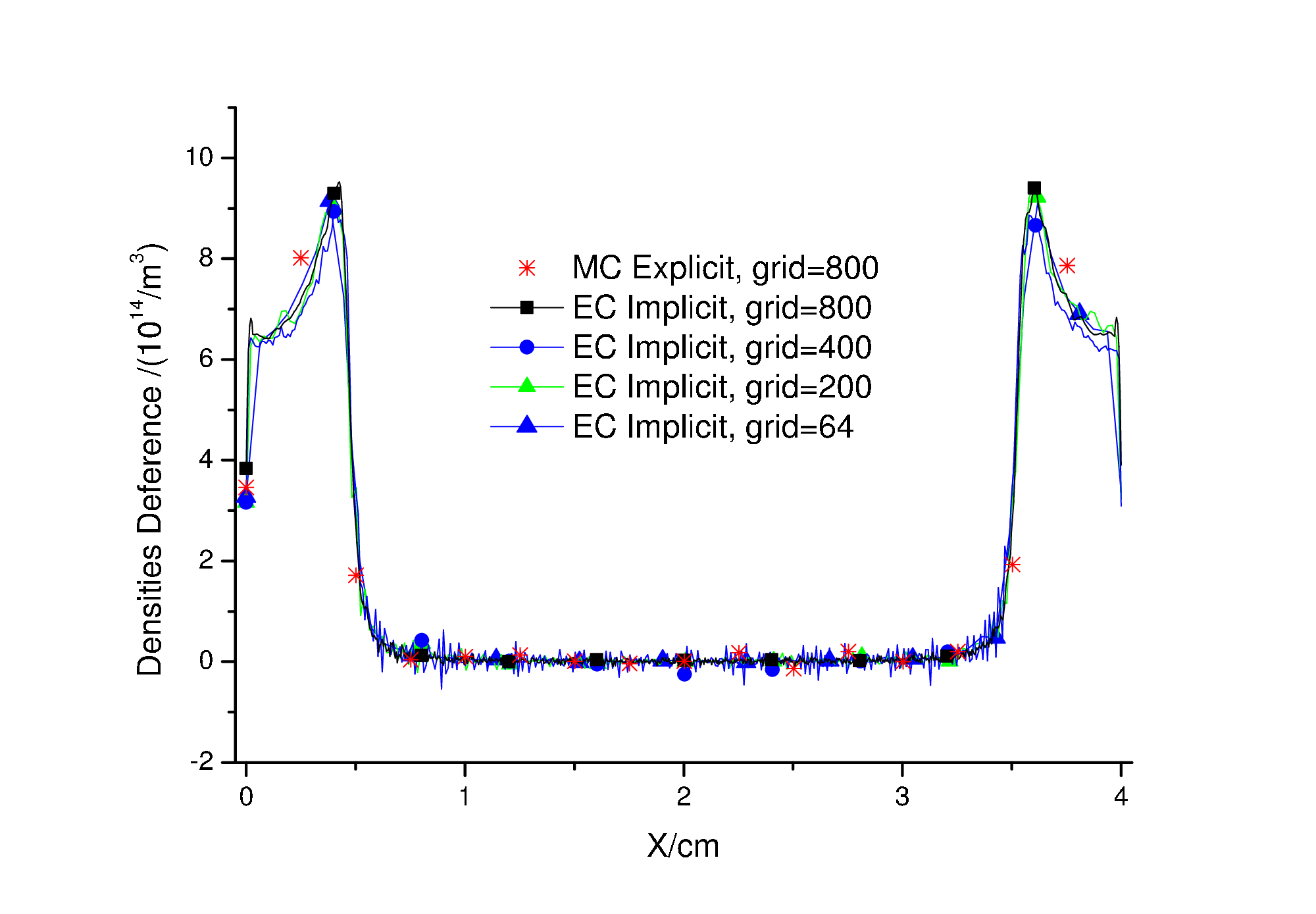}
\caption{The $\delta n=n_i-n_e$ show the sheath structure, left for Explicit, right for Implicit }
\label{sheath}
\end{figure}

\subsubsection{Heating Rates}
The electron heating rates with stable discharging are shown in figure \ref{hr}. It can be found that the heating rates of EC scheme are lower and the peaks are more narrow than the MC scheme.  We can understand the phenomena by considering the
sheath heating mechanism{\cite{Lieberman05}}: the electron collides with the moving sheath and is accelerated. This can be considered as a electron being bounced with
an oscillating sheath. After being bounced back from the oscillating sheath, it runs towards the bulk plasma and the other electrode. The accelerating  behavior
depends on the phase matching of the plasma sheath oscillating and the electron oscillating. On the other hand, EC scheme causes the self force of particles. The self force gives the particle an slight and fast oscillating\cite{Langdon73}.  Then the EC scheme causes orbit error slightly, the error will accumulate and disturb the accelerating. Hence the phase matching lengths are shorter and the heating rates peaks are more narrow. The implicit
algorithms works better than the explicit algorithms: when grid=64, the heating peaks height (comes from scholastic heating) reduces to about 70\% in EC Explicit scheme. In contrast, the peaks height keeps on almost same value in EC Implicit scheme grid=64. We think the possible reason is that the implicit algorithms damp the high frequency oscillation which includes the
self force disturbing, then the accelerating phase is more accurate. In addition, the higher accuracy of heating rates should be the origin of more precise density and voltage.

As a comparison, we showed the heating rates of MC implicit scheme with $grid=400$ and same $dt$. In fact, under these parameters, the MC implicit simulating can give out converge results but with higher densities. When the grid spacing becomes wilder, the MC implicit scheme diverges like the
MC explicit scheme. We can find the reason in the heating rates profile. The scale of this case is showed on the right. We can find the peak heating rates in this case are total wrong(perhaps because of the wrong densities). In addition, we find a bulk heating region in the middle of the discharging area. The larger the grid spacing becomes, the stronger the bulk heating becomes. Even if the MC scheme can give out some converge results, it is qualitatively wrong when $DX > 3\lambda_D$, which is the major strongpoint of the EC scheme.
\begin{figure}

\includegraphics[scale=0.4]{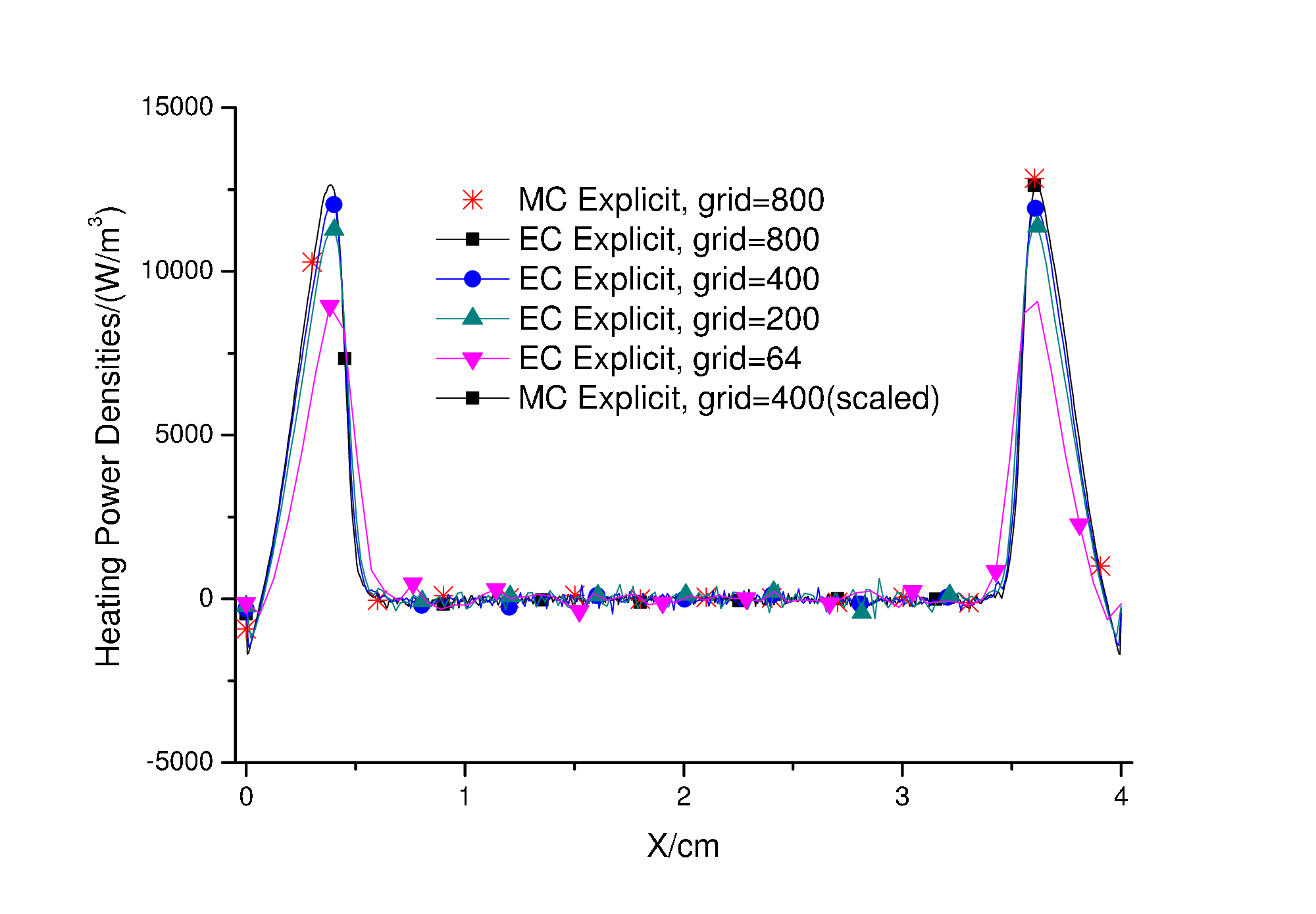}
\includegraphics[scale=0.4]{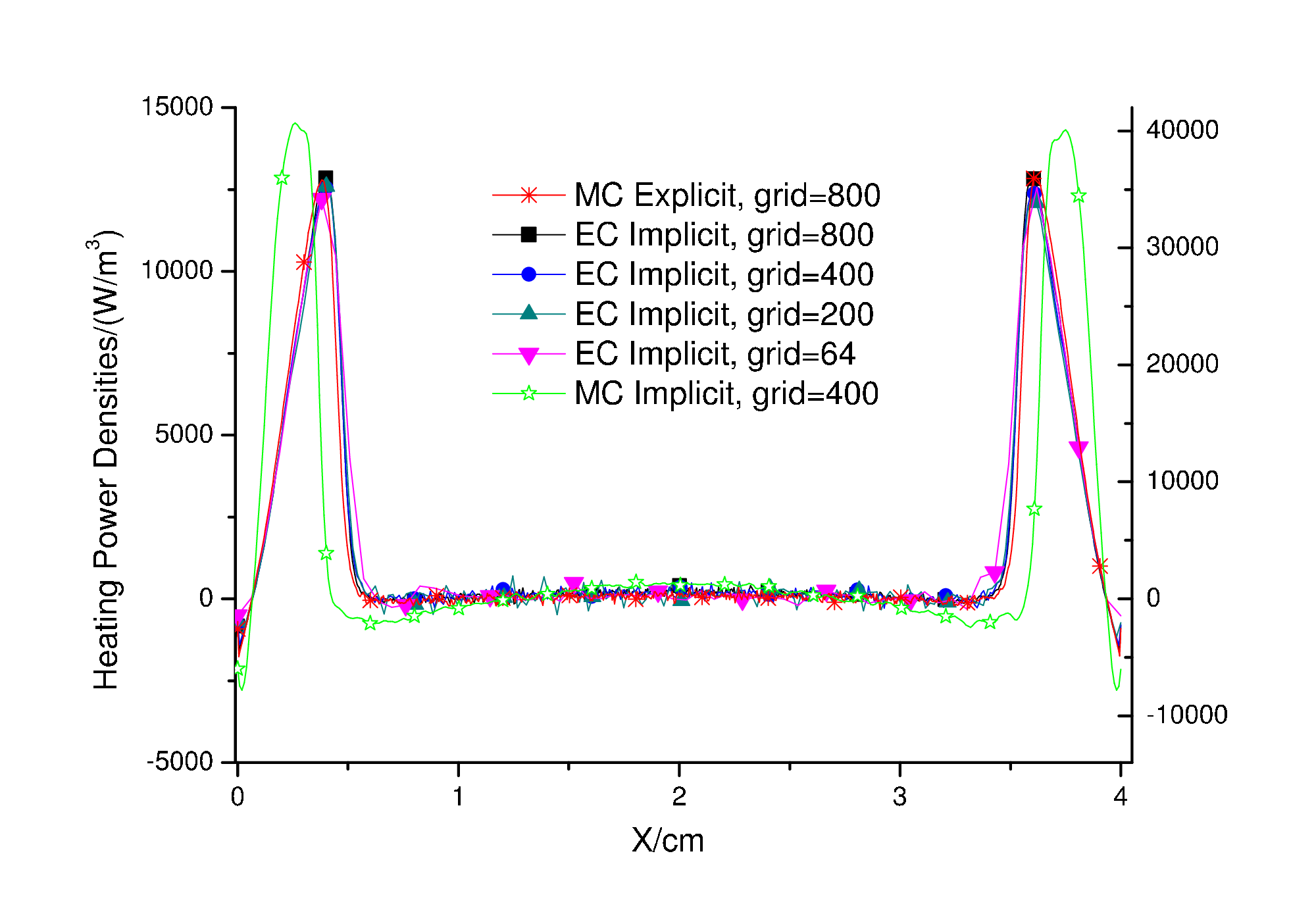}
\caption{Electron heating rates while stable discharging.  Left for Explicit, right for Implicit. In the right figure,
the left axis scale is for the
results of the benchmark and the EC scheme. The result with MC implicit (green line) is re-scaled to fit in the figure and can be read with the right axis scale. One can find the heating rates of MC implicit grid 400 are much larger than other and have peak in the middle.}
\label{hr}
\end{figure}

\subsubsection{Energy Distribution}
The electron energy distribution is shown in firgue {\ref{espec}}. At 50 mTorr pressure, the energy distribution of electrons is a slight  bi-Maxwellian distribution.The results of the EC explicit simulation are more close to single temperature Maxwellian, especially when the grid spacing becomes large. In addition, the electron's average energy in EC explicit scheme becomes slightly higher(about $5\%-10\%$) than MC benchmark.
This effects come from the particle stochastic noise increasing(for the particle number decreasing) partially and from the essentiality of the EC scheme partially.  In theory, the particle self force in EC scheme causes more "momentum exchanging" between the particle and the grid and can also induce more energy re-distribute among the electrons. Therefore the results are not unexpected.

The behavior of the EC implicit scheme is slightly more complex. In the EC explicit scheme, the grid spacing increasing causes the Electron Energy Probability Function (EEPF) be thermalized. However, the direct implicit scheme causes some damping and stronger bi-Maxwellian. Accordingly, we can see some lines are below the benchmark and some others are above it.  Anyway, for the very large grid spacing, the EC implicit will be always better than EC explicit.

For the analysis, we draw out the result of the MC implicit scheme while the grid spacing is larger than $\xi\lambda_D$. The distribution of electron energy has very long tail which indicates self-heating. This is in agreement with our judgement: the MC scheme causes qualitatively wrong results while $dx > \xi\lambda_D$.
\begin{figure}
\includegraphics[scale=0.4]{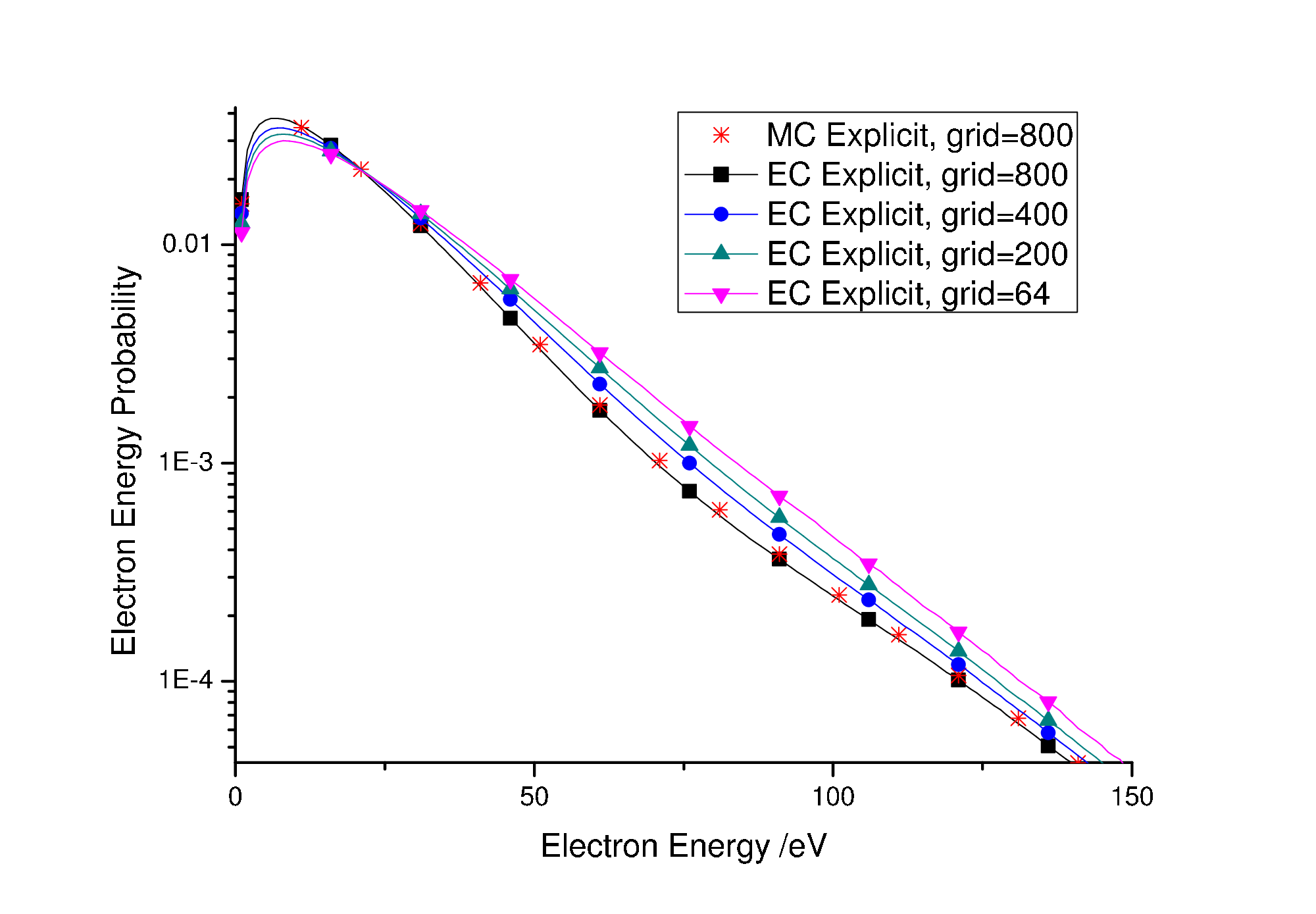}
\includegraphics[scale=0.4]{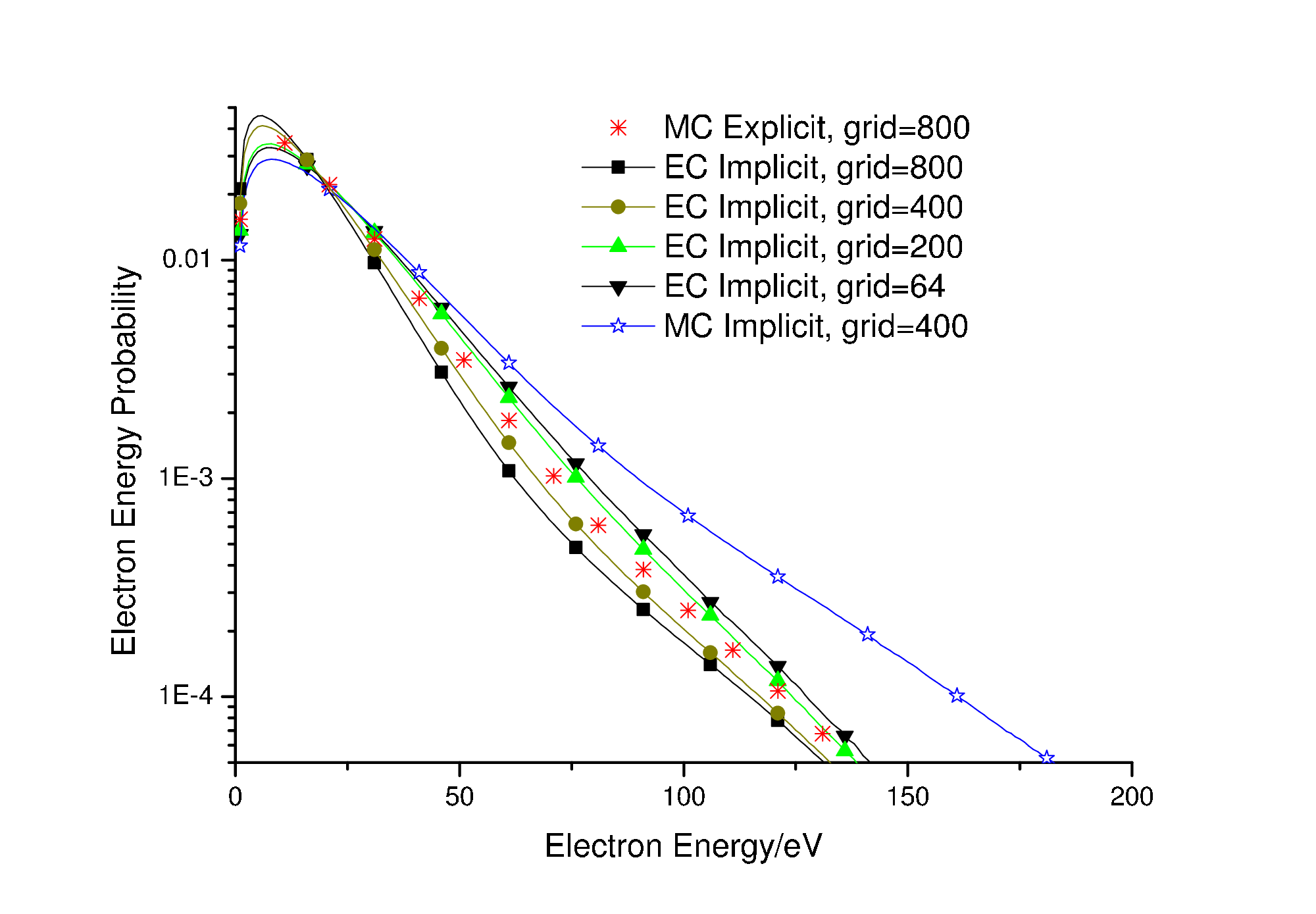}
\caption{The EEPF of the electrons. Left for Explicit, right for Implicit. When the grid increasing, the electron's distribution becomes single temperature Maxwellian. The results of MC Implicit grid 400 (green line) is shown in right figure also. the very long tail showed self (grid) heating.}
\label{espec}
\end{figure}

For the convenience of the analysis, we show the $\langle \frac 12 mv_x^2\rangle$ in figure \ref{enervx}. In 1D simulating, the electron
bouncing and accelerating are occurring in direction X and transfer the kinetic energy to other directions by collision.
Thus the equivalence kinetic energy  $\langle \frac 12 mv_x^2\rangle$ can indicate the bouncing and accelerating. One can find
the distribution of the equivalence kinetic energy in X direction is a standard two temperature maxwellian distribution. When
 the grid spacing increases, the
temperature of high energy part increases slightly in the EC scheme. This increasing is caused by the decreased particle number partially and
by the EC self-force disturbing partially. The MC implicit results are shown in the figure also. There are longer tail in it but not as
evidential as the energy distribution figure. It agrees with the judgement that the grid heating took place in all directions and was not
a resonance effect.

\begin{figure}
\includegraphics[scale=0.4]{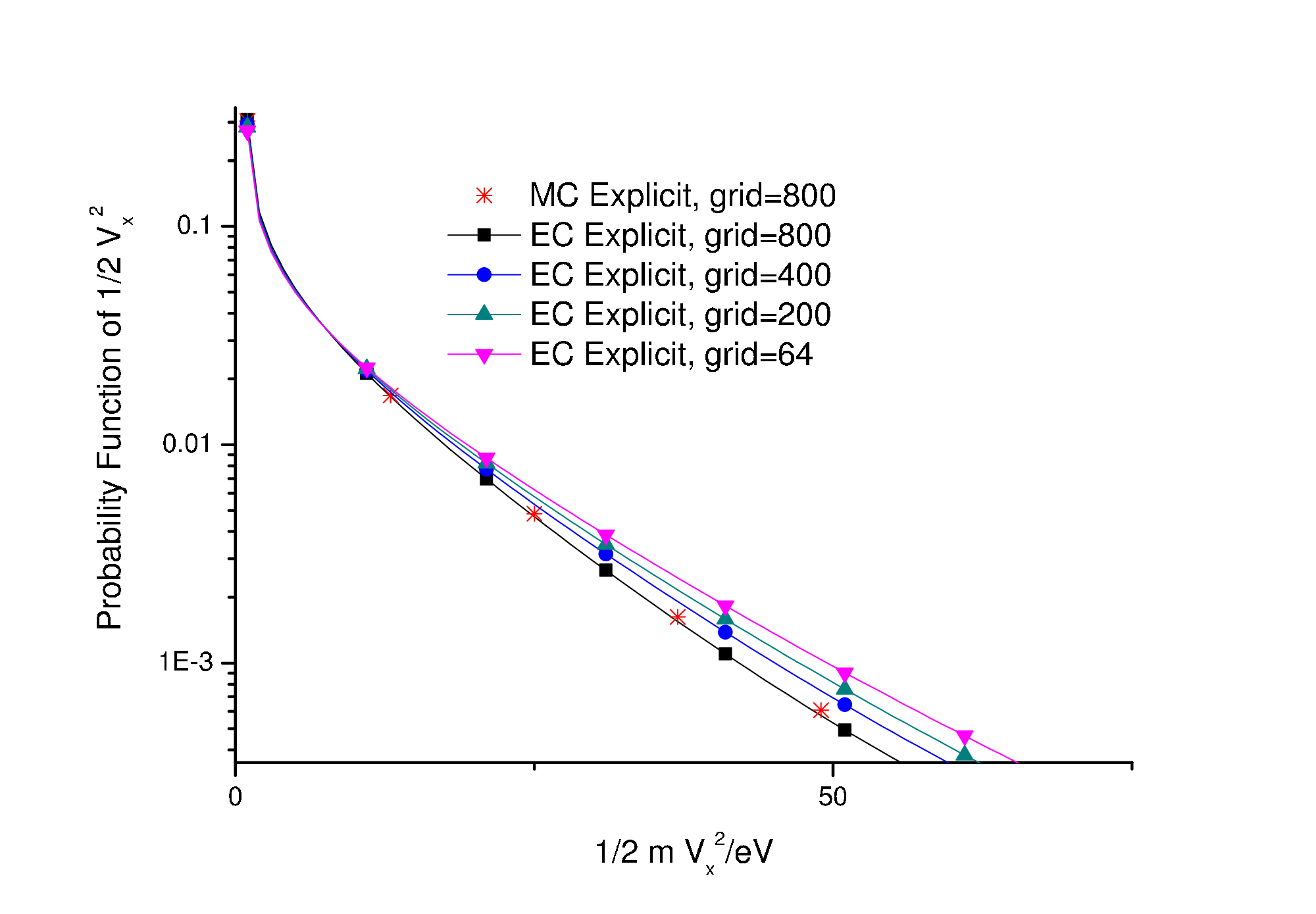}
\includegraphics[scale=0.4]{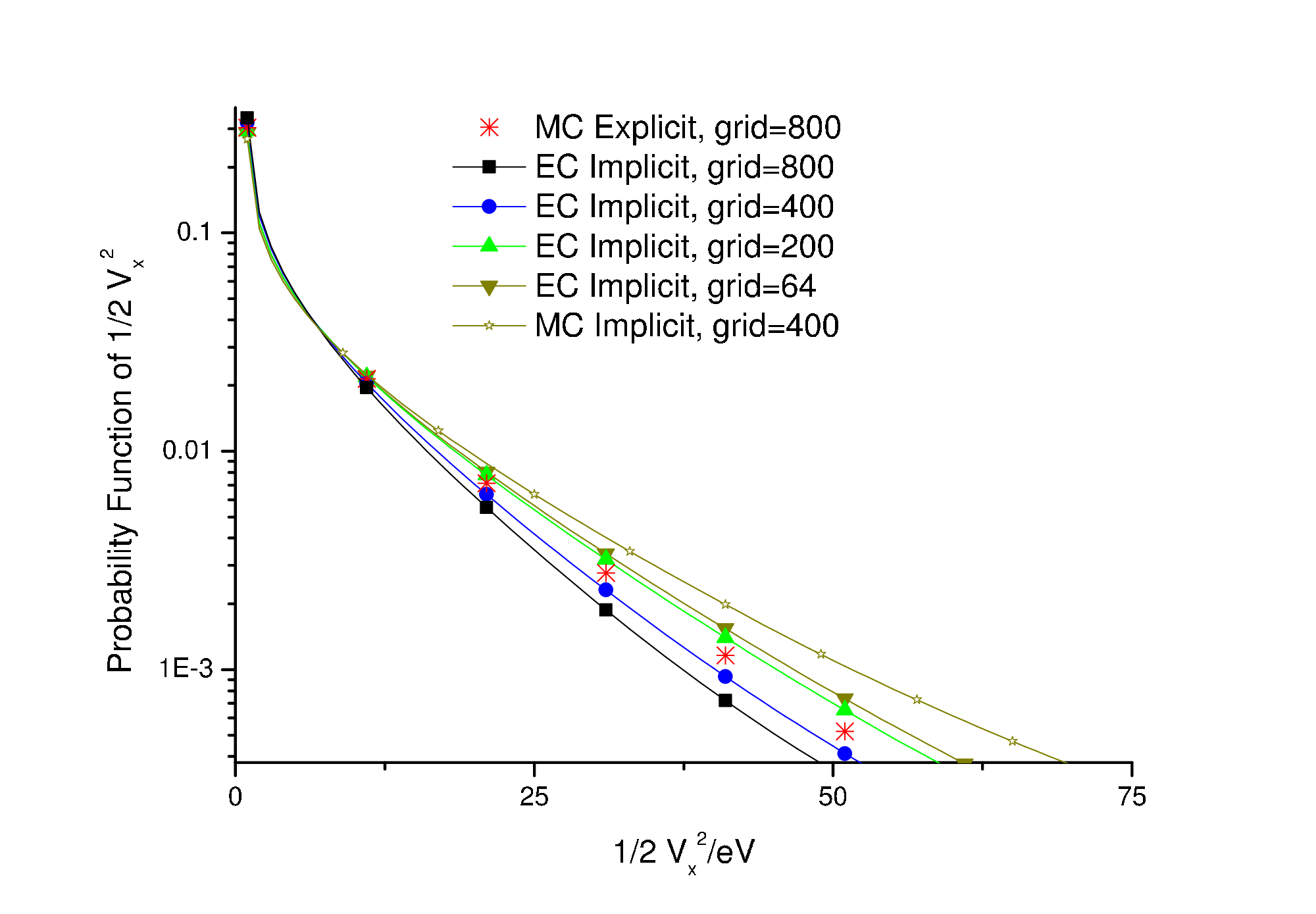}
\caption{The distribution of the electrons equivalence kinetic energy in direct X. Left for Explicit, right for Implicit. When the grid increasing, the high energy part increase the
temperature slightly.}
\label{enervx}
\end{figure}

\subsection{The effects of time steps}
Theoretically, the explicit PIC algorithm will diverge while $dt>2./\omega_p$ and the implicit algorithm can give out stable results.
However, because of the grid heating and the Monte Carlo Couple of the particles, the results could change. We show the average electron
densities and heating rates with $dt=1.5*10^{-11}$ to $dt=1.0*10^{-10}$ in  figure \ref{edenvsdt} and figure  \ref{heatvsdt}. One can find the
relations between dt and the average electron densities are slightly complex. The reason can be found in figure  \ref{heatvsdt}. As we expected,
the increase of the dt causes the decrease of the peaks of the stochastic heating in implicit algorithms,
while this increase affects little on the peaks in the explicit algorithms. The effects can be explained by the well known "self-cooling" effects of
direct implicit PIC algorithms. However, some bulk heating results appear in the large dt simulations. The bulk heating seems to increase more
in implicit algorithms.
We are not very clear about its source. We think they are some pseudo Ohmic heating. That is, when the dt becomes large,
the average acceleration in one step is slightly larger than the real field acceleration for the error of the leap-frog scheme, which enables some pseudo Ohmic heating to occur. In the implicit algorithms, the algorithms decrease the high frequency distorts of fields (decrease the collision damping) then make the situation even worse.  In fact, when the $dt>2./\omega_p$, we find that even the implicit algorithms can not give out converge results. Anyway, the pseudo Ohmic heating is small when the $dt<0.5/\omega_p$ ($<5\%$ of the scholastic heating)and show no heating peak in all cases.

\begin{figure}
\includegraphics[scale=0.5]{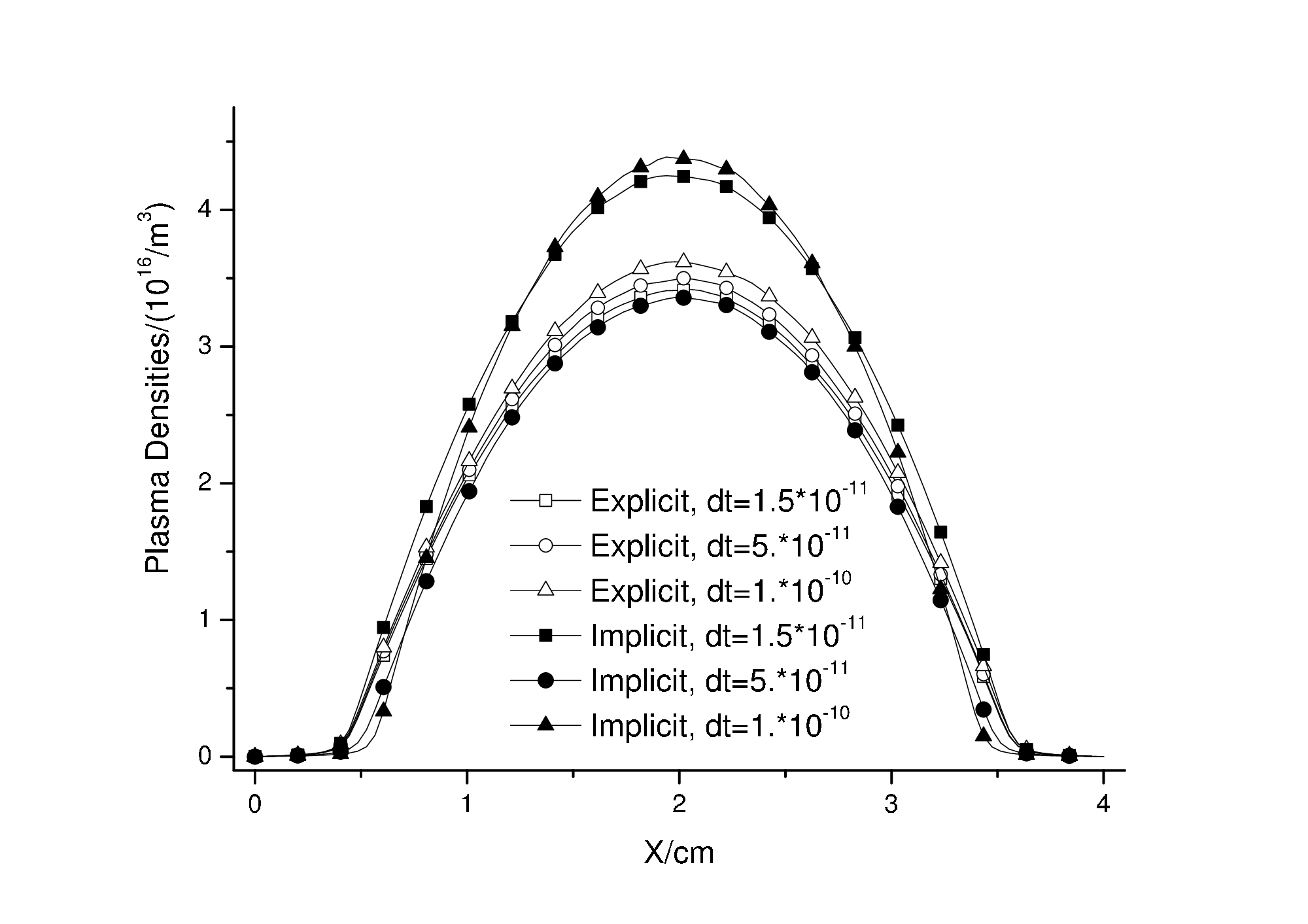}
\caption{The average electron
densities vs dt, The grid numbers are set to 100.}
\label{edenvsdt}
\end{figure}

\begin{figure}
\includegraphics[scale=0.4]{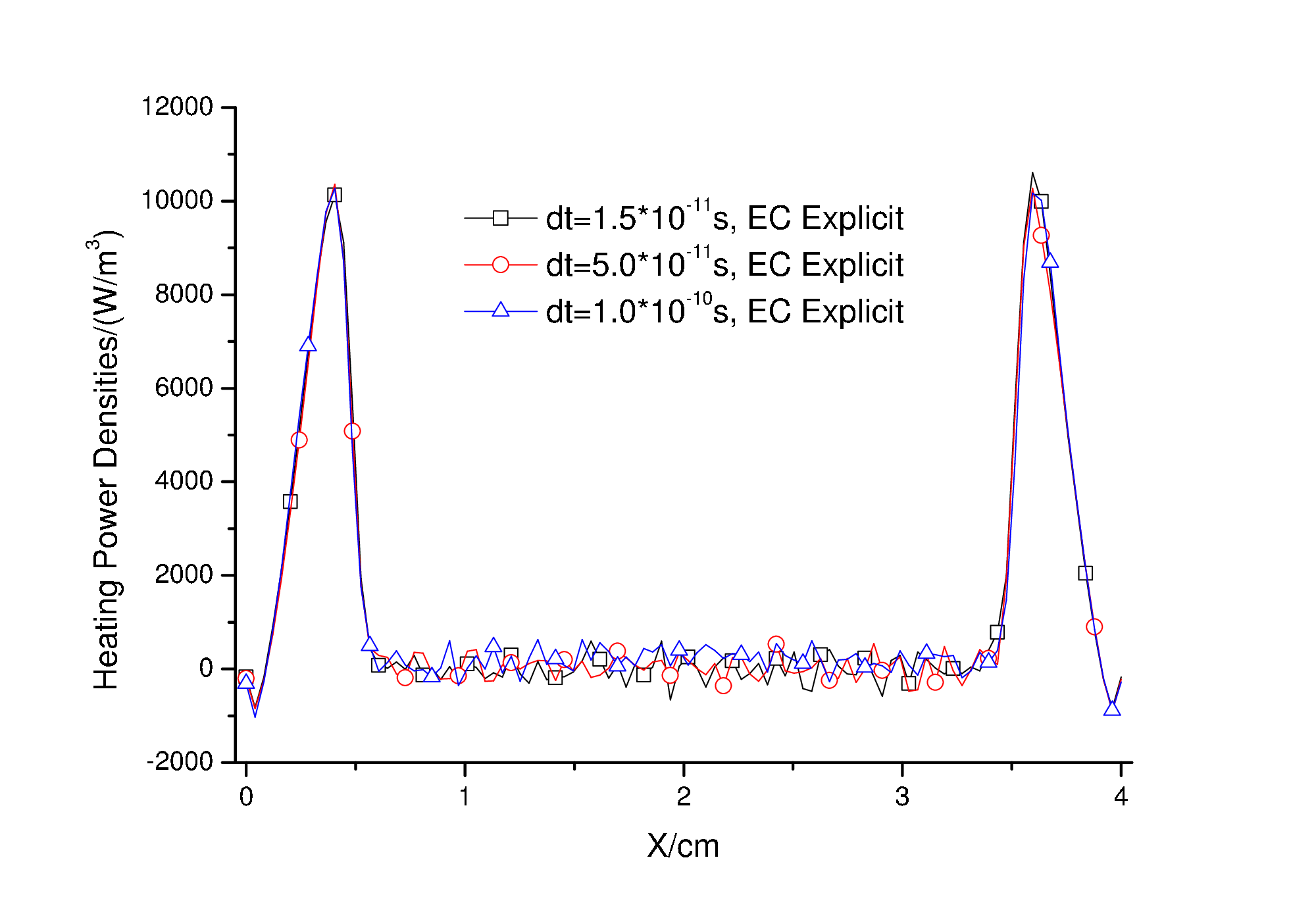}
\includegraphics[scale=0.4]{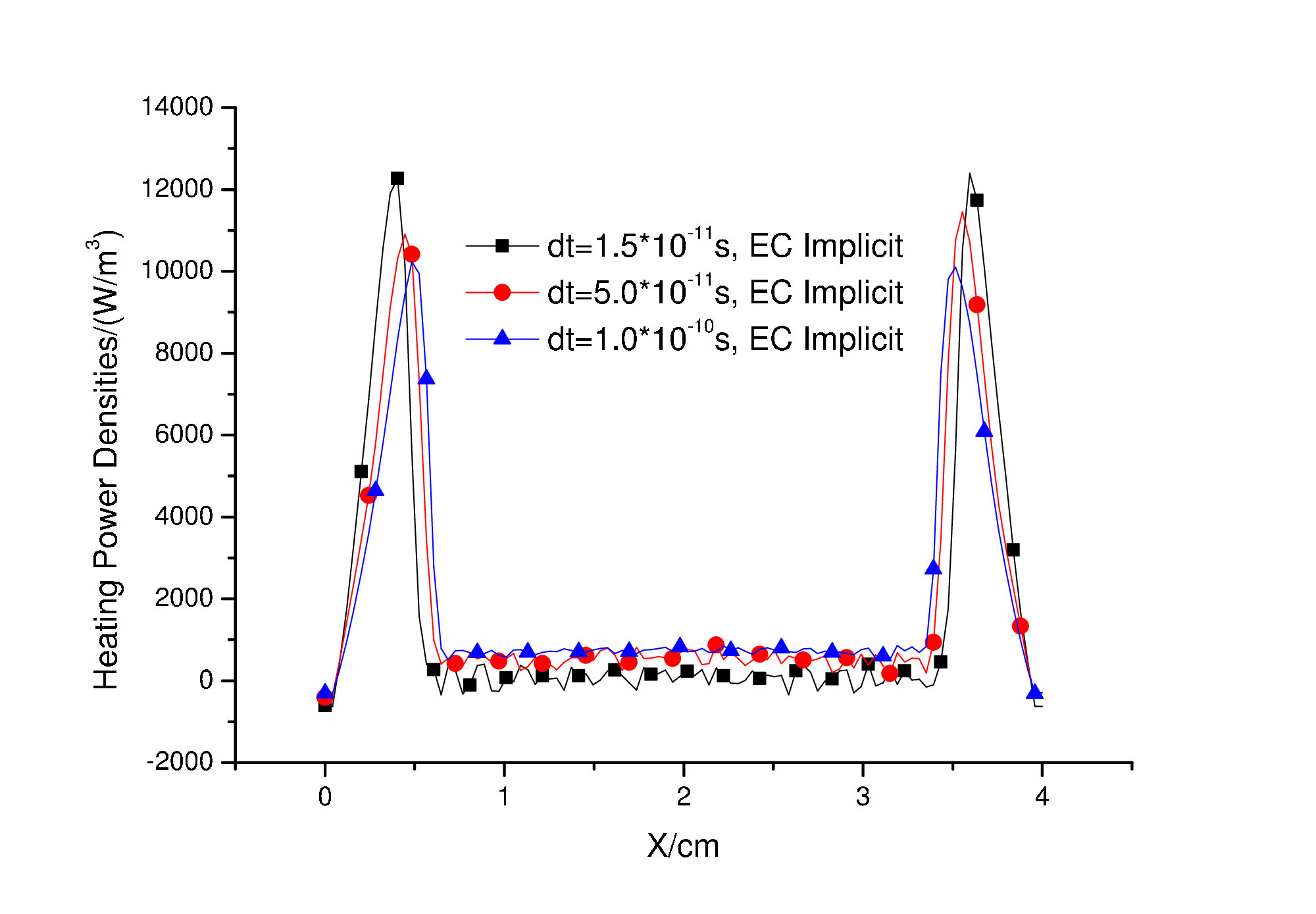}
\caption{The average heating rates
vs dt, left for Explicit, right for Implicit. The grid numbers are set to 100.}
\label{heatvsdt}
\end{figure}

\subsection{The effects of macro particle numbers per cell(PPC)}

As Tunner showed {\cite{Tunner06}}, the PIC/MCC simulating results can be affected by macro particle per Debye length($N_D$) and total macro particle numbers($N_T$).  In our problems,the effects can be shown with the results by PPC. This effect comes from stochastic noise and can cause simulating breakdown if $N_D$ is too small. We showed the effects of PPC with EC implicit and EC explicit while grid number is 100 and 200. The results are more complex than the upper effects. The PPC effects can be found easily, and implicit schemes work better than explicit scheme always. However, in the case of implicit scheme, the PPC effects can overrun the grid number effects. The reason could be that the grid number effects have been reduced much in implicit scheme.
\begin{figure}
\includegraphics[scale=0.5]{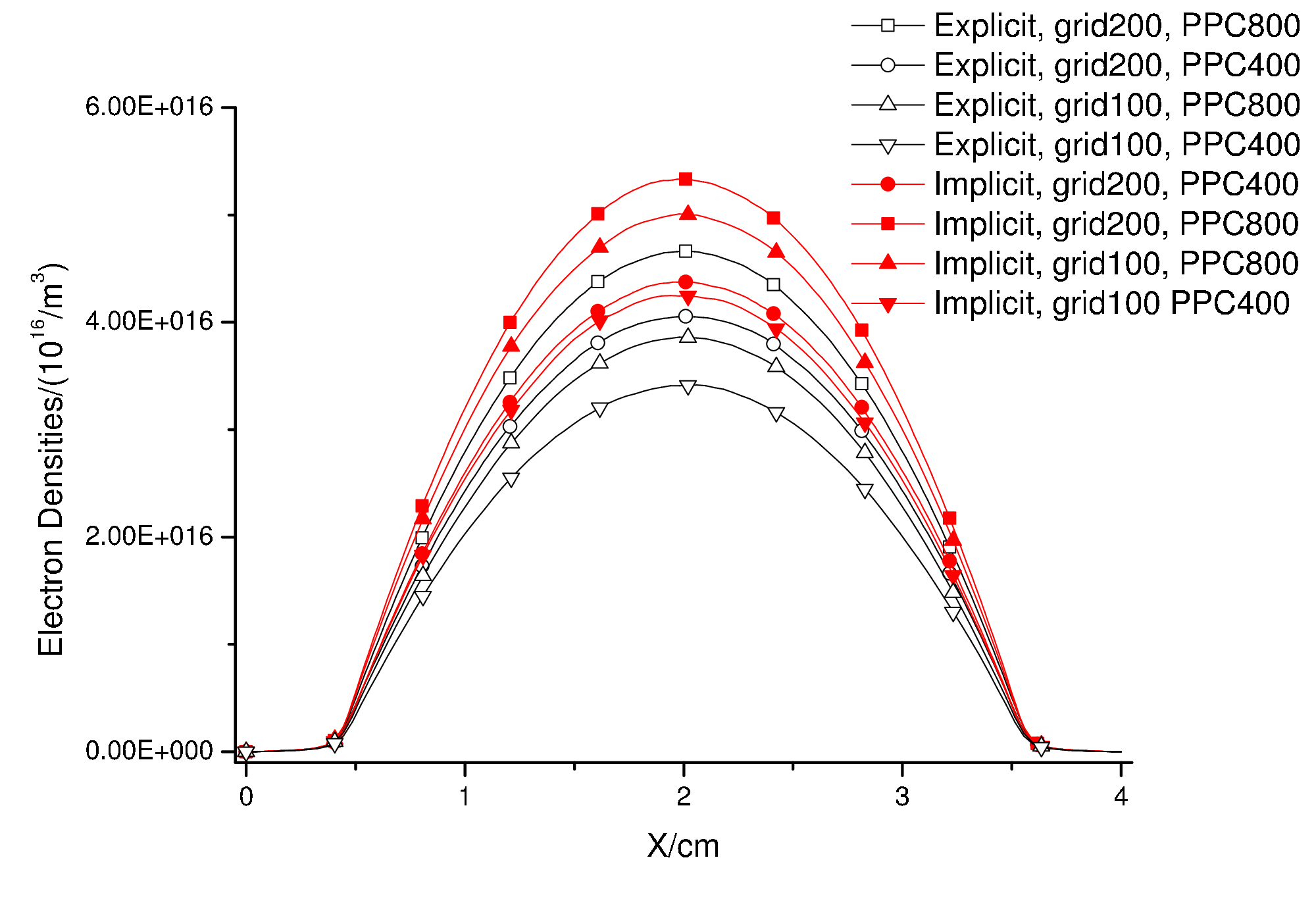}
\caption{The electron densities by PPC=400 to 800 while grid number is 100 and 200.}
\label{denvsppc}
\end{figure}

\section{Discussion}

In the above simulations, we compared the results from convenient momentum conservation PIC scheme and
energy conservation PIC scheme. The EC scheme had shown it can control self-heating behavior and can converge with much more wider grid. However, while the grid spacing becomes very wide, numerical cooling-like effects will appear obviously. Fortunately, the important physics is still preserved. In our simulation, Not only the diffusion behaviors like sheath formation are always correct, but also the kinetics effects as power absorption are conceptually correct.In addition, the combination of implicit and EC scheme showed higher accuracy than conventional explicit EC scheme.

A contrast result is the MC implicit scheme. Although the implicit scheme introduces numerical damping and counteracts the self heating partially, the self heating causes diverge in many cases. In addition, the heating rates showed the MC implicit scheme can give out qualitative  wrong
results in some worst cases.

We can conclude that the combination of EC scheme and direct implicit algorithm is optimum numerical scheme for the  PIC/MCC simulations,
especially very large space and time steps in this scenario can be adopted, so that the computational costs can be significantly reduced. At the same time , the physical properties of the system are preserved well and numerical instability can be avoided, if the space and time steps are not too large. While in conventional MC scheme, the numerical error may lead to non-physical heating of the plasma, because the energy is not exactly conserved.

However, this scheme also has some shortcomings. Because the momentum is not conserved, the energy distributions of the electrons will be slightly disturbed by the self force when the particles pass the cell boundaries. There is no effective technology to overcome this problem at present. Fortunately, the disturbing is small in most cases. The major properties of the system like the plasma density and temperature are still simulated well.

 We also showed that some other non-physical effects of the particles in EC scheme will occur at very large space and time steps. The results indicate that the error of the stochastic heating rates comes from the small oscillation of EC scheme. Because the essential oscillation is controlled by the grid spacing, there is no perfect method to get rid of the error. However, our results indicate that the error is only important in the sheath heating region and implicit algorithms migrate them much. For the gas discharge simulations, we could apply a multi-scale method to overcome the problem. That is, we can use much finer grid in the sheath region and use wide grid in the other region. For the other problem (for example, ECR), more powerful technologies are needed. If the problems are not dominated by the stochastic heating and the orbit resonance, EC scheme could give more accuracy results and could be applied credibly.

The final problem is the pulse discharge. In these kinds of discharge, the heating rates are important, especial their profile. We have shown the EC
implicit scheme can give out correct stochastic heating rate profile. In addition, we can expect the resonance heating and Ohmic heating can be qualitatively correct in the scheme. When the $dt$ becomes large, a subtle problem shows that some non-physical heating (perhaps pseudo Ohmic heating) appears, but it is not very large. We expect that the shortcoming will not prevent its applying in the pulse discharge simulating.


\begin{thebibliography}{99}
\itemsep=-4pt plus.2pt minus.2pt  
\small
\bibitem{Birdsall91} Birdsall C K and Langdon A B 1985  {\it Plasma Physics via Computer
Simulation } (McGraw-Hill: New York)
\bibitem{Verboncoeur05}Verboncoeur J P 2005 {\it Plasma Phys. Control. Fusion} \textbf{47} A231
\bibitem{Lieberman05}Lieberman M A and Lichtenberg A J 2005 {\it Principles of
    Plasma Discharges and Materials Processing} (2nd edn) (Wiley: New York)
\bibitem{Zhao08} Zhao H Y and Mu Z X 2008 {\it Chin Phys} B  \textbf{17} 1475
\bibitem{Shi09} Shi F, Zhang L L and Wang D Z 2009 {\it Chin Phys} B  \textbf{18} 1674
\bibitem{Liu10}Liu C S, Han H Y, Peng X Q, Chang Y and Wang D Z 2010 {\it Chin Phys} B \textbf{19} 035201
\bibitem{Jin09}Jin X L, Huang T, Liao P, Yang Z H 2009 {\it Acta Phys. Sin.} \textbf{58} 5526  (in Chinese)
\bibitem{Wang13}Wang H H, Liu D G, Meng L,  Liu L Q, Yang C, Peng K and Xia M Z 2013 {\it Acta Phys. Sin.} \textbf{62} 015207 (in Chinese)
\bibitem{Tskhakaya07} Tskhakaya D, Matyash K , Schneider R and Taccogna F 2007, {\it Contrib. Plasma. Phys.} \textbf{47} 563
\bibitem{Hockney88} Hockney R W and Eastwood J W 1988 {\it Computer Simulation Using Particles} (Adams Hilger: New York)


\bibitem{Taccogn04} Taccogn F, Longo S, Capitelli M and Schneider R 2004  {\it Comput. Phys. Commun.} \textbf{164}  160
\bibitem{Abe86} Abe H, Sakairi N and Itatani R 1986 {\it J. Comput. Phys} \textbf{63} 247
\bibitem{Sentoku08} Y. Sentoku, A. J. Kemp, J. Comput. Phys. 227, 6846 (2008)
\bibitem{Cai10} Cai H B, Mima K ,Sunahara A , Johzaki T, Nagatomo H, Zhu S P and He X T 2010 {\it  Phys. Plasma} \textbf{17} 023106
\bibitem{Langdon83} Langdon A B, Cohen B I and Friedman A 1983 {\it J. Comput. Phys} \textbf{51} 107
\bibitem{Friedman90}Friedman A 1990 {\it  J. Comput. Phys.} \textbf{90} 292
\bibitem{Wang10}Wang H Y, Jiang W and  Wang Y N 2010 {\it Plasma Source Sci. Technol.} \textbf{19} 045023
\bibitem{Vahedi93} Vahedi V, DiPeso G, Birdsall C K, Lieberman M A and Rognlien T D 1993 {\it Plasma. Source. Sci. Technol.} \textbf{2} 261
\bibitem{Ricci02}Ricci P, Lapenta G and Brackbill J U  2002 {\it J. Comput. Phys.} \textbf{183} 117
\bibitem{Welch06}Welch D R, Rose D V , Cuneo M E, Campbell R B and Mehlhorn T A 2006  {\it Phys. Plasma} \textbf{13} 063105
\bibitem{Eastwood91}Eastwood J W 1991 {\it Comput. Phys. Commun.} \textbf{64} 252
\bibitem{Eastwood95}Eastwood J W, Arter W , Brealey N J, Hockney R W 1995 {\it Comput. Phys. Commun.} \textbf{87} 155
\bibitem{Pointon08} Pointon T D 2008 {\it Comput. Phys. Commun.} \textbf{179} 535
\bibitem{Markidis11} Markidis S and Lapenta G 2011 {\it Journal of Computational Physics} \textbf{230} 7037
\bibitem{Brackbill85} Brackbill J U and Cohen B I 1985 {\it Multiple Time Scales} (Academic Press Inc: London)
\bibitem{Birdsall91b} Birdsall C K 1991 {\it IEEE. Trans. Plasma. Sci.} \textbf{19} 65
\bibitem{Langdon73} Langdon A B 1973 {\it J. Comput. Phys.} \textbf{12} 247
\bibitem{Jiang08}Jiang W, Xu X, Dai Z L and Wang Y N 2008 {\it Phys. Plasma.} \textbf{15} 033502
\bibitem{Xu92}Xu H X and Brackbill J U 1992 {\it Comput. Phys. Commun.} \textbf{69} 253
\bibitem{Tunner06} Turner M M 2006 {\it Phys. Plasma.} \textbf{13} 033506
\end{thebibliography}
\end{document}